\documentclass[11pt,a4paper]{article}
\pdfoutput=1
\usepackage{exscale,times}
\usepackage{fancyhdr}
\usepackage[cp1250]{inputenc}
\usepackage[T1]{fontenc}
\usepackage{amsmath}
\usepackage{fancyhdr}
\usepackage{indentfirst}
\usepackage{times}
\usepackage{graphicx}
\usepackage{multicol}
\usepackage{here}
\usepackage{amsmath}
\usepackage{amssymb}
\usepackage{amsthm}
\usepackage{amsfonts}
\usepackage{stmaryrd}
\usepackage{mathrsfs}
\usepackage{color}
\usepackage{multicol}
\usepackage{mathrsfs}
\usepackage{wasysym}
\usepackage[yyyymmdd,hhmmss]{datetime}

\newcommand{\bal}{\mbox{\boldmath $\alpha$}}

\newcommand{\bup}{\mbox{\boldmath $\upsilon$}}

\newcommand{\bsigma}{\mbox{\boldmath $\sigma$}}

\newcommand{\bveps}{\mbox{\boldmath $\varepsilon$}}

\newtheorem{myRemark}{Remark}

\newcommand{\mytitle}{Fractional Calculus for Continuum Mechanics - anisotropic non-locality}

  \usepackage{hyperref}
 \hypersetup{
      bookmarks=true,         % show bookmarks bar?
      pdftoolbar=true,        % show Acrobat?s toolbar?
     pdftitle={\mytitle},    % title
      pdfauthor={Wojciech Sumelka},     % author  
      colorlinks=true,       % false: boxed links; true: colored links
      linkcolor=blue,          % color of internal links (change box color with linkbordercolor)
      citecolor=blue,        % color of links to bibliography
  	bookmarksnumbered=true
 }

\begin{document}
\pagestyle{fancy}

\chead{Wojciech Sumelka - Compiled on \today\ at \currenttime}
\lhead{}
\rhead{}
\renewcommand{\headrulewidth}{0.4pt}
%
%   Please insert the title here...
%
\begin{center}
{\fontsize{14}{20}\bf{\mytitle}}
\end{center}

\begin{center}
\textbf{W. Sumelka}\\
\bigskip
{Poznań University of Technology, Institute of Structural Engineering, }\\
Piotrowo 5 street, 60-969 Poznań, Poland\\ 
wojciech.sumelka@put.poznan.pl\\

\bigskip
\end{center}

%%%%%%%%%%%%%%%%%%%%%%%%%%%%%%%%%%%%%%%%%%%%%%%%%%%%%%%%%%%%%%%%%%%%%%%
%\begin{document}
%\pagestyle{fancy}
%
%\chead{Journal of Elasticity}
%\lhead{}
%\rhead{}
%\renewcommand{\headrulewidth}{0.4pt}
%
%\begin{center}
%%
%%   Please insert the title here...
%%
%{\fontsize{14}{20}\bf{Non-locality of Fractional Deformation}}
%\end{center}
%
%\begin{center}
%\textbf{Wojciech Sumelka}\\
%\bigskip
%{Poznan University of Technology, Institute of Structural Engineering, }\\
%Piotrowo 5 street, 60-969 Poznan, Poland\\ 
%wojciech.sumelka@put.poznan.pl\\
%\bigskip
%\end{center}
%
%   Please insert the keywords here...
%


\vspace{1cm}

\begin{center}
\textbf{ABSTRACT}\\[1mm]
\end{center}
In this paper the generalisation of previous author's formulation of fractional continuum mechanics to the case of \textit{anisotropic non-locality} is presented. The considerations include the review of competitive formulations available in literature. The overall concept bases on the \textit{fractional deformation gradient} which is non-local, as a consequence of fractional derivative definition. The main advantage of the proposed formulation is its analogical structure to the general framework of classical continuum mechanics. In this sense, it allows, to give similar physical and geometrical meaning of introduced objects.

\section{Introduction}

Reliable modelling of heterogeneous materials, ranging from marco- to micro- and nanoscale of observation, in terms of continuum mechanics concept, needs non-local formulations  \cite{Pamin2004}. The first articles in this area were released in the 1960s, and dates back to the scientists, such as Toupin \cite{Toupin1962}, Mindlin \cite{Mindlin1965}, Eringen \cite{Eringen-1967}, Dillon \cite{Dillon-1970},  Dafalias \cite{Dafalias-1977}, Ba$\mathrm{\check{z}}$ant  \cite{Bazant-1979}, Maugin \cite{Maugin1979}, Aifantis \cite{Aifantis1992}, Fleck and Hutchinson \cite{Fleck97-AAM}. 
Regardless of the details of specific formulation, it is common that non-local model introduces characteristic length (or time) which is inherent to the inner material structure. This new material parameter can be directly measured from experiment or can be determined from experiments via inverse analysis or derived theoretically from micromechanics \cite{Geers-1997,Cuenot-2000,Yang-2002,Lam-2003,Ji-2010,Liebold-2013}.

Nowadays there are many concepts dealing with non-local formulations, like {general non-local theories} \cite{Peddieson2003,Eringen2010}, {strain-gradient theories} \cite{Toupin1962,Mindlin1968}, {micropolar theories} \cite{Eringen1966,Nowacki1972} or {theories of material surfaces} \cite{Gurtin1975}. Nevertheless, due to development of new materials, and constant miniaturization of e.g. electronic or medical devices, the continued progress in this subject is desirable. In some sense, as a response to this demand the new branch of investigations was initiated in early 2000's by Klimek \cite{Klimek2001}, Vazquez \cite{Vazaquez2004}, Lazopoulos \cite{Lazopoulos2006}, namely the one which utilises fractional calculus.

{An important class on non-local models are those dealing with anisotropic non-locality - especially the one combined with damage description. Such formulations introduce the possibility of analysis of materials showing scale effect but depending on direction. Let us mention herein, the papers in this subject by: Kuhl et al. \cite{Kuhl2000} where quasi-brittle materials where considered; Stumpf et al. \cite{Stumpf2001} where the crack analysis based on the concept of continua with microstructure and evolving defects was discussed; Germain et al. \cite{Germain2007} where the anisotropic layered material was formulated; Abu-Al-Rub et al. \cite{Abu-Al-Rub2009} were coupled anisotropic damage and plasticity constitutive model to predict the concrete distinct
behavior in tension and compression was considered; Alastrue et al. \cite{Alastrue2009} where fully three-dimensional anisotropic elastic model for vascular tissue modelling was shown; and finally by Perzyna \cite{Perzyna2008} and Sumelka et al. \cite{Sumelka2010b-JTAM2010,Sumelka2012-CM} where class of implicitly non-local (rate type \cite{Needleman1988}) anisotropic models for metallic materials were considered. Nevertheless from the point of view of subject of this paper, the fractional anisotropic non-local models are nowadays still under development.}
 
In this paper we propose a generalisation of an original concept of \textit{isotropic} fractional continuum mechanics presented in \cite{Sumelka2013-work2013,Sumelka2013-TH} for description of anisotropic non-locality. As will be discussed this formulation abandon not-only classical postulate of local action, but also restriction imposed by objectivity postulate (cf. discussion on objectivity in \cite{Frewer-2009}). In this sense, this result will be analogous to the one obtained by Drapaca and Sivaloganathan \cite{Drapaca2012}, but is should be emphasised that both models can not be reduced one to each other, and operate in different physical dimension space.

As mentioned the subject of this paper, namely the definition of non-local model utilising fractional calculus, belongs to the recent trends in mechanics \cite{Klimek2001,Vazaquez2004,Lazopoulos2006,Paola2009,Atanackovic2009,Carpinteri2011,Drapaca2012}. In comparison to previous formulations the main advantage of the proposed formulation is its analogical structure to the general framework of classical continuum mechanics. Therefore, it allows, to some extent, to give similar physical and geometrical meaning of introduced objects according to their classical counterparts (including of course some open questions due to interpretation of fractional calculus itself \cite{Podlubny2002}). Other crucial advantages are: (i) we deal with finite deformations; (ii) the generalised fractional measures of the deformation e.g. fractional deformation gradients or fractional strains have the same physical dimensions as classical one; (iii) characteristic length scale of the particular material is defined explicitly.

The paper is structured as follows. In Section 2 existing fractional models of mechanics are summarised. The non-local fractional model accounting for anisotropic non-locality, is presented in Section 3. In Section 4, a benchmark examples are shown.

\begin{myRemark}\textbf{Naming convention.} 
Throughout the paper, we follow the naming convention common for the classical continuum mechanics.
To show the influence of fractional calculus, we add the word 'fractional'. Therefore, we introduce objects such as
fractional deformation gradient or fractional strain.

It is important to notice, that similar nomenclature (because of fractional calculus application) is utilized in theories dealing with fractal media, where fractal/fractional continua is considered \cite{Carpinteri1997,Carpinteri2002,Tarasov2005a,Tarasov2005b,OstojaM-2007,Drapaca2012,Balankin2013}.
Those two concepts should not be confused. We read the influence of fractional differential operator, as in presented formulation, from phenomenological point of view leaving other interpretations for the future work. In other words, we examine some smoothed picture, without going into detail about the sub-scale constituents/phenomena.
\end{myRemark}

\section{Existing fractional models of mechanics}

Let us shortly discuss selected concepts of non-local fractional continuum mechanics.

Non-locality in the fractional continuum mechanics comes from the application of fractional derivative which operates over the interval. This interval defines simultaneously the range of non-local interaction. The fractional derivative means the derivative of an arbitrary order \cite{Podlubny1999-AP,Kilbas06}, and in a special case when the applied order becomes integer we obtain classical local formulation (classical local derivative where the definition is given in a single point). In this sense, and because there are many different fractional differential operators \cite{Oliveira-2014}, and many different ways how to introduce them into the specific formulation, we have nowadays several fractional continuum mechanics concepts. In general, one can distinguish two ways of introducing spatial non-local effects through fractional calculus into continuum mechanics \cite{Atanackovic-2014-Willey}: (i) redefinition of kinematics; (ii) redefinition of constitutive law.

One of the first papers dealing with fractional kinematics was by Klimek \cite{Klimek2001} where the symmetric fractional derivative had been used in the definition of strain, namely
\begin{equation}\label{eq:klimek}
\mathcal{K}^\alpha(x,t)=\frac{1}{2}\left( ~_{0}D^{\alpha}_x -~_{x}D^{\alpha}_L  \right)u(x,t),
\end{equation}
where $\mathcal{K}^\alpha$ is a fractional strain, $D^\alpha$ is a fractional derivative, $x$ is a spatial coordinate, $t$ denotes time, and $u$ is a displacement. In is important that the definition by Eq.~(\ref{eq:klimek}) describes one dimensional problem under small strain assumption, the terminals $0$ and $L$ show that for calculation of fractional strain at the specific point of interest $x$ we take all information from the body, and the fractional strain has physical unit $[m^{1-\alpha}]$. Another proposition was by Lazopoulos \cite{Lazopoulos2006}. He proposed a measure of fractional strain in a form
\begin{equation}\label{eq:Lazopoulos}
\mathcal{L}^\alpha(x,t)=\frac{\partial u(x,t)}{\partial x}\left( ~_{0}D^{\alpha}_x +~_{x}D^{\alpha}_L  \right)\frac{\partial }{\partial x}u(x,t),
\end{equation}
where $\mathcal{L}^\alpha$ stands for fractional strain, with similar comments as for Eq.~(\ref{eq:klimek}). Later a closely related proposition was stated by Atanackovic and Stankovic \cite{Atanackovic2009}, namely
\begin{equation}\label{eq:Atanackovic}
{\cal{E}}^\alpha(x,t)=\frac{1}{2}\left(~_{-\infty}^C D^{\alpha}_x- ~_{x}^C D^{\alpha}_\infty\right)u(x,t),
\end{equation}
where $\cal{E}^\alpha$ stands for fractional strain.

Seminal work dealing with the proposition of fractional kinematics at finite strain for three dimensions was the paper by Drapaca and Sivaloganathan \cite{Drapaca2012}. They started from the redefinition of motion, namely
\begin{equation}\label{eq:Drapaca}
\mathbf{x}=
\begin{cases}
\tilde{\mathbf{K}}_{\mathbf{X}}^{\mathbf{1}-\bal(t)}\chi(\mathbf{X},t),\quad\quad -\infty<\alpha_{Ii}(t)<1,\quad I,i=1,2,3,\\
\chi(\mathbf{X},t),\quad\quad\quad\quad\quad \alpha_{Ii}(t)=1,\quad I,i=1,2,3,
\end{cases}
\end{equation}
where $\chi$ is a motion, $\mathbf{x}$ is a spatial coordinate, $\mathbf{X}$ denotes a material coordinate, {and $\bal$ is an order of motion}. In Eq.~(\ref{eq:Drapaca}) 
\begin{equation}
\tilde{\mathbf{K}}_{\mathbf{X}}^{\mathbf{1}-\bal(t)}=
\begin{bmatrix}
\tilde{\mathbf{K}}_{\mathbf{X}}^{\mathbf{1}-\bal_{\mathbf{1}}(t)} & 0 & 0\\
0 & \tilde{\mathbf{K}}_{\mathbf{X}}^{\mathbf{1}-\bal_{\mathbf{2}}(t)} & 0\\
0 & 0 & \tilde{\mathbf{K}}_{\mathbf{X}}^{\mathbf{1}-\bal_{\mathbf{3}}(t)}\\
\end{bmatrix},
\end{equation}
{where $\bal_{\mathbf{i}}=(\alpha_{1i},\alpha_{2i},\alpha_{3i})$, so}
\begin{equation}
\bal=
\begin{bmatrix}
\alpha_{11} & \alpha_{12} & \alpha_{13} \\
\alpha_{21} & \alpha_{22} & \alpha_{23} \\
\alpha_{31} & \alpha_{32} & \alpha_{33} \\
\end{bmatrix},
\end{equation}
and finally
{
\begin{eqnarray}
\nonumber
x_i&=&\tilde{\mathbf{K}}_{\mathbf{X}}^{\mathbf{1}-\bal_{\mathbf{i}}(t)} \chi_i=\tilde{R}^{1-\alpha_{1i}}_{X_1} \tilde{R}^{1-\alpha_{2i}}_{X_2} \tilde{R}^{1-\alpha_{3i}}_{X_3} \chi_i=\\
~&~&= \frac{1}{\Gamma(1-\alpha_{1i}(t))\Gamma(1-\alpha_{2i}(t))\Gamma(1-\alpha_{3i}(t))}\\
\nonumber
~&~&\int\int\int_H \frac{\chi_i(\mathbf{Y},t)dY_1dY_2dY_3}{|X_1-Y_1|^{\alpha_{1i}(t)}|X_2-Y_2|^{\alpha_{2i}(t)}|X_3-Y_3|^{\alpha_{3i}}(t)},
\end{eqnarray}}
where $H=[a_1,b_1]\times[a_2,b_2]\times[a_3,b_3]$ denotes the interval of non-local interaction, {and $\Gamma$ is the Euler gamma function
\begin{equation}
\Gamma(\alpha)=\int_{0}^{\infty}e^{-t}t^{\alpha-1}dt.
\end{equation}}Next, based on non-standard definition of motion by Eq.~(\ref{eq:Drapaca}) the authors define the deformation gradient $\mathbf{F}^{\bal}$ as 
\begin{equation}
\mathbf{F}^{\bal}=\frac{\partial \chi(\mathbf{X},t)}{\partial \mathbf{X}}, \quad \mathrm{or}\quad F^{\alpha}_{aA}=\frac{\partial}{\partial X_A}\left( \tilde{R}^{1-\alpha_{1a}}_{X_1} \tilde{R}^{1-\alpha_{2a}}_{X_2} \tilde{R}^{1-\alpha_{3a}}_{X_3} \chi_a \right) \mathbf{e}_a\varotimes\mathbf{E}_A,
\end{equation}
{where $\mathbf{e}_a~\textrm{and}~\mathbf{E}_A$ are base vectors in coordinate systems $\{x_a\}~\textrm{and}~\{X_A\}$, respectively.}

Finally, the operator $\mathbf{F}^{\bal}$, which is now non-local, plays the same role as the classical deformation gradient. It is important to point out that $\mathbf{F}^{\bal}$ has a physical unit, in this sense fractional strains also, and also for the first time the non-local fractional kinematics is defined on an interval which represents a part of the body, and enables to analyse anisotropic non-local effect.

Concerning the second group of non-local fractional models, namely the one which bases on redefinition of constitutive law, let us first mention about paper by Di Paola et al. \cite{Paola2009}. In this paper the non-local elasticity relation was proposed as
\begin{equation}\label{eq:Paola}
\sigma(x,t)-\ell^\alpha {\cal{E}}_x^\alpha\sigma(x,t)=E\varepsilon(x,t),
\end{equation}
where $\sigma$ is a Cauchy stress, $\ell^\alpha$ is a length scale, ${\cal{E}}_x^\alpha$ is a fractional derivative, and $E$ denotes Young modulus. It is clear that Eq.~(\ref{eq:Paola}) is a generalisation of well known Eringen model \cite{Eringen-1983}. Later,
Carpinteri et al. \cite{Carpinteri2011} considered a material constitutive law in a form
\begin{equation}\label{eq:Carpinteri}
\sigma(x,t)=E\left[ \frac{du}{dx}+\frac{\kappa_\alpha}{2} \left(~_{a}^C D^{\alpha}_x u- ~_{x}^C D^{\alpha}_b u\right)\right],
\end{equation}
where $\kappa_\alpha$ is a material constant and has anomalous physical dimensions. Interestingly, Eq.~(\ref{eq:Carpinteri}) reflects the same basic formula as the one derived from Atanackovic and Stankovic \cite{Atanackovic2009} but starting from fractional
non-local strain measure \cite{Carpinteri2014}.

As a concluding remark, it should be emphasised that an interesting direction of research is also an attempt to unify non-local fractional models with theories dealing with fractal media or peridynamic model \cite{Silling2000}.

\section{Fractional kinematics, stresses and balance laws}

\subsection{Riesz-Caputo fractional derivative}

There are many definitions of fractional derivative \cite{Samko93,Podlubny1999-AP,Kilbas06,Leszczynski2011,Oliveira-2014}. They all share one common attribute i.e. they are all defined on an interval contrary to integer order differential operators defined in a point. In following part of this paper the Caputo's type derivative over the interval $(a,b)$ is considered. We call such operator Riesz-Caputo (RC) derivative cf. \cite{Frederico2010} and its definition includes linear combination of left and right Caputo's derivatives, namely
\begin{equation}\label{eq:RC}
~_{~~a}^{RC}D^{\alpha}_b f(t)=\frac{\varsigma(\alpha)}{2}\left(~_{a}^{C}D^{\alpha}_t f(t)+(-1)^n~_{t}^{C}D^{\alpha}_b  f(t)\right),
\end{equation}
where $\varsigma(\alpha)$ represents scalar valued function, $~_{a}^{C}D^{\alpha}_t f(t)$ is a left-sided Caputo's derivative given by ($t>a$ and $n=[\alpha]+1$)
\begin{equation}\label{eq:CaL}
~_{a}^{C}D^{\alpha}_t f(t)=\frac{1}{\Gamma(n-\alpha)}\int_a^t \frac{f^{(n)}(\tau)}{(t-\tau)^{\alpha-n+1}}d\tau,
\end{equation}
$~_{t}^{C}D^{\alpha}_b  f(t)$ is a right-sided Caputo derivative given by ($t<b$ and $n=[\alpha]+1$)
\begin{equation}\label{eq:CaR}
~_{t}^{C}D^{\alpha}_b f(t)=\frac{(-1)^n}{\Gamma(n-\alpha)}\int_t^b \frac{f^{(n)}(\tau)}{(\tau-t)^{\alpha-n+1}}d\tau.
\end{equation}

In the remaining part of this paper the RC derivative defined in Eq.~(\ref{eq:RC}) is shortly denoted as $D^{\alpha}$ with the possibility of writing variable under the $D$ in case of a partial differentiation of multivariate functions. For example $\underset{X_1}{D} ~^\alpha f $ represents partial fractional derivative of $f$ with respect to the variable $X_1$ over the interval which should be explicitly defined before $X_1\in(a,b)$. 
It is clear that for $\alpha=1$ we have
\begin{equation}
~_{~~a}^{RC}D^{1}_b f(t)=\frac{\mathrm{d}}{\mathrm{d}t}f(t).
\end{equation}

\begin{myRemark}\textbf{Caputo's derivative of a constant function.} In general fractional derivative of a particular type of a constant function is \textbf{not} equal zero. Caputo's derivative is an exception, thus makes its application similar to classical integer order operator.
\end{myRemark}

\begin{myRemark}\textbf{On initial and boundary conditions.}
It is fundamental that using Caputo's type derivative, one requires standard (like in the classical differential equations) initial and$/$or boundary conditions, whereas for other types of fractional derivatives (e.g. RL or GL) the initial/boundary conditions are of different type dependently on chosen definition.
\end{myRemark}

\begin{myRemark}\textbf{On the type of applied fractional differential operator.}
The type of applied fractional differential operator describes the type of non-locality. In other words, it governs the way in which the information from the surrounding influences particular point of interest. Thus, one should choose appropriate fractional differential operator, dependently on the material considered (in contrast the choice of RC in the presented paper is due to reasons stated in Remarks 2 and 3). 

Recall, as an example, that the definition of the classical Riesz-Feller \cite{Feller52} fractional operator has an origin in processes with Lévy stable probability distribution. In this sense it should be possible to define fractional differential operator in a way that e.g. it has an information about the distribution of grains sizes in a particular metal.

\end{myRemark}

\subsection{Fractional deformation gradients}

It is important to emphasise that the specific fractional model presented in this paper was discussed in a series of papers devoted to fractional elasticity \cite{Sumelka2014-AoM,Sumelka2014-AAM}; fractional thermoelasticity \cite{Sumelka2013-TH}, fractional Kirchhoff-Love Plates \cite{Sumelka2014-ACME}, and non-local rate independent plasticity \cite{Sumelka2014-AM}. Nevertheless, these results comprise the case of \textit{isotropic} non-locality and it was shown that for such a case objectivity restriction known from classical mechanics holds.

As mentioned in the introduction, in this paper we analyse the case of \textit{anisotropic} non-locality i.e. materials showing scale effect but depending on direction. Therefore, this formulation abandon not-only classical postulate of local action, but also restriction imposed by objectivity postulate - what can be important  for some classes of materials \cite{Frewer-2009}. In this sense, this result is analogous to the one obtained by Drapaca and Sivaloganathan \cite{Drapaca2012}, but is should be emphasised that both models can not be reduced one to each other, and operate in different physical dimension space. 

The description is given in the Euclidean space. We refer to $\mathcal{B}$ as the reference configuration of the continuum body while $\mathcal{S}$ denotes its current configuration. Points in $\mathcal{B}$ are denoted by $\mathbf{X}$ and in $\mathcal{S}$ by $\mathbf{x}$. Coordinate system for $\mathcal{B}$ is denoted by $\left\{X_{A}\right \}$ with base $\mathbf{E}_{A}$ and for $\mathcal{S}$ we have $\left\{x_{a}\right \}$ with base $\mathbf{e}_{a}$. 

The regular motion of the material body $\mathcal{B}$ can be written as 
\begin{equation}\label{eq:motion}
\mathbf{x}=\phi(\mathbf{X},t),
\end{equation}
and its inverse as
\begin{equation}
\mathbf{X}=\varphi(\mathbf{x},t),
\end{equation}
thus $\phi_{t}:\mathcal{B}\shortrightarrow \mathcal{S}$ is a $C^{1}$ actual configuration of $\mathcal{B}$ in $\mathcal{S}$, at time $t$. Here, it is important to emphasise that to avoid confusion we introduce new symbol for motion marked in Sec. 2 by $\chi$ where we have followed Drapaca and Sivaloganathan \cite{Drapaca2012} notation. In this sense, in the presented formulation the position of the non-local body is not changed contrary to \cite{Drapaca2012}, but the effort of the body (strains/stresses) is of course different due to non-local action.

We define the fractional deformation gradient and its inverse as follows ($\alpha \in (0,1)$)
\begin{equation}\label{eq:fracdefGradX}
\underset{X}{\tilde{\mathbf{F}}}(\mathbf{X},t)=\underset{X}{\ell}^{\bal-1}\underset{X}{D}^{\bal} \phi(\mathbf{X},t),\quad \mathrm{or}\quad {\underset{X}{\tilde{F}}} ~_{aA}=\underset{X}{\ell} ~_{aA}^{\alpha_{aA}-1}\underset{X_A}{D}^{\alpha_{aA}}\phi_a\mathbf{e}_a\varotimes\mathbf{E}_A,
\end{equation}
and
\begin{equation}\label{eq:fracdefGradx}
\underset{x}{\tilde{\mathbf{F}}}(\mathbf{x},t)=\underset{x}{\ell}^{\bal-1}\underset{x}{D}^{\bal}\varphi(\mathbf{x},t),\quad \mathrm{or}\quad {\underset{x}{\tilde{F}}} ~_{Aa}=\underset{x}{\ell} ~_{Aa}^{\alpha_{Aa}-1}\underset{x_a}{D}^{\alpha_{Aa}}\varphi_A\mathbf{E}_A\varotimes\mathbf{e}_a,
\end{equation}
where ${D}^\alpha$ is a fractional differential operator in the sense of RC defined in previous section, and $\ell_X$ and $\ell_x$ are length scales in $\mathcal{B}$ and $\mathcal{S}$, respectively. 
As an example the matrix representation of object $\underset{X}{\tilde{\mathbf{F}}}$ is
\begin{equation}
\underset{X}{\tilde{\mathbf{F}}}=
\begin{bmatrix}
\underset{X}{\ell} ~_{11}^{\alpha_{11}-1}\underset{X_1}{D}^{\alpha_{11}}\phi_1 & \underset{X}{\ell} ~_{12}^{\alpha_{12}-1}\underset{X_2}{D}^{\alpha_{12}}\phi_1 & \underset{X}{\ell} ~_{13}^{\alpha_{13}-1}\underset{X_3}{D}^{\alpha_{13}}\phi_1 \\
\underset{X}{\ell} ~_{21}^{\alpha_{21}-1}\underset{X_1}{D}^{\alpha_{21}}\phi_2 & \underset{X}{\ell} ~_{22}^{\alpha_{22}-1}\underset{X_2}{D}^{\alpha_{22}}\phi_2 & \underset{X}{\ell} ~_{23}^{\alpha_{23}-1}\underset{X_3}{D}^{\alpha_{23}}\phi_2 \\
\underset{X}{\ell} ~_{31}^{\alpha_{31}-1}\underset{X_1}{D}^{\alpha_{31}}\phi_3 & \underset{X}{\ell} ~_{32}^{\alpha_{32}-1}\underset{X_2}{D}^{\alpha_{32}}\phi_3 & \underset{X}{\ell} ~_{33}^{\alpha_{33}-1}\underset{X_3}{D}^{\alpha_{33}}\phi_3 
\end{bmatrix}.
\end{equation}

Hence, in general we can write

\begin{equation}
\underset{x}{\tilde{\mathbf{F}}}\underset{X}{\tilde{\mathbf{F}}}\neq \mathbf{I}=\delta_{AB}\mathbf{E}_A\varotimes\mathbf{E}_B,
\end{equation}

and 

\begin{equation}
\underset{X}{\tilde{\mathbf{F}}}\underset{x}{\tilde{\mathbf{F}}}\neq \mathbf{i}=\delta_{ab}\mathbf{e}_a\varotimes\mathbf{e}_b,
\end{equation}
{where $\delta_{ij}$ denotes the Kronecker delta.}

\begin{figure}[H]
\centering
\includegraphics[width=10cm]{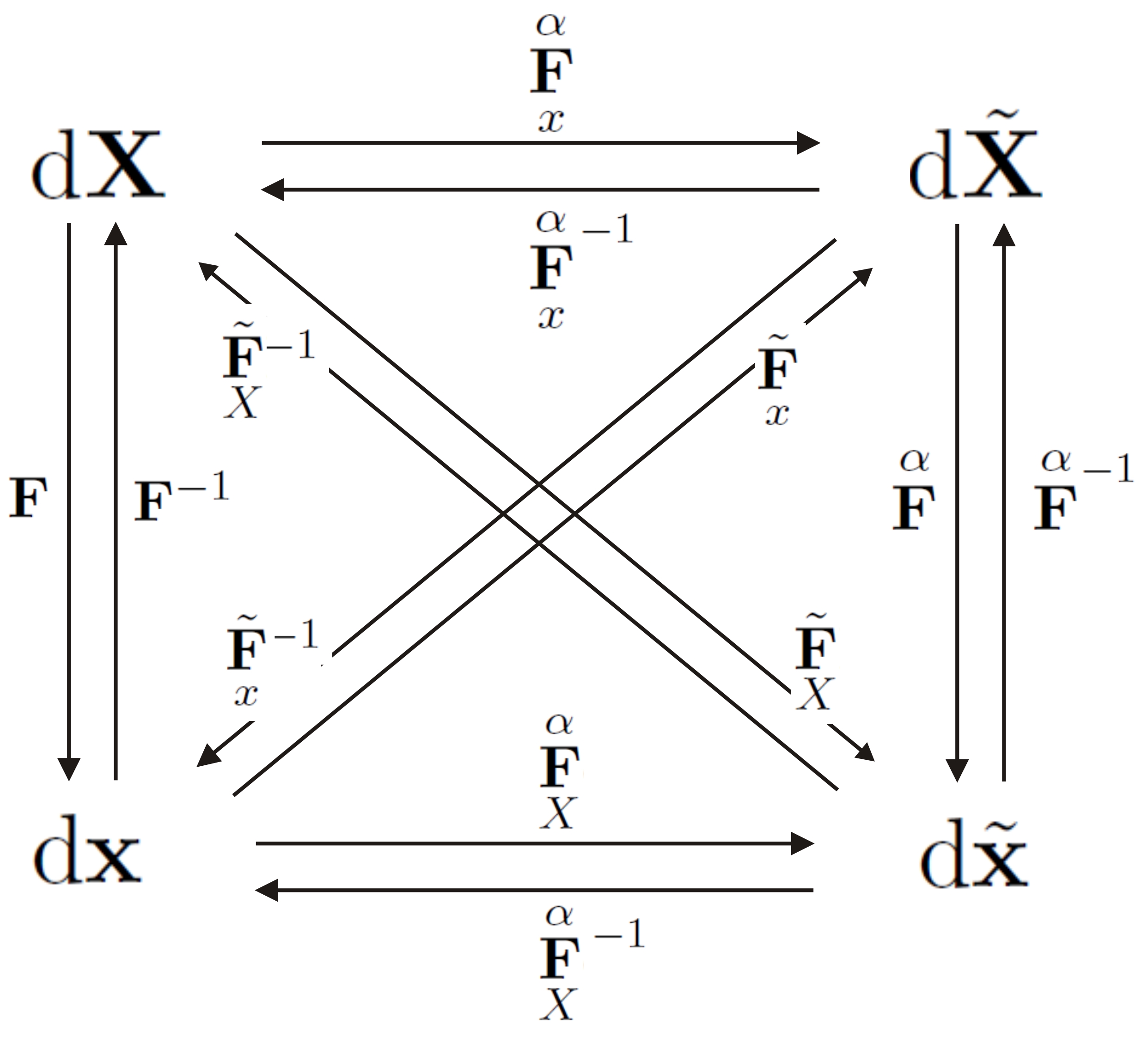}
\caption{The relations between material and spatial line elements with their fractional counterparts}
\label{fig:diagram}
\end{figure}

Having in mind classical definition of material $\mathrm{d}\mathbf{X}$ and spatial $\mathrm{d}\mathbf{x}$ line elements, namely
\begin{equation}\label{eq:dx}
\mathrm{d}\mathbf{x}=\mathbf{F}\mathrm{d}\mathbf{X},\quad \mathrm{or}\quad \mathrm{d}x_a=F_{aA}\mathrm{d}X_A\mathbf{e}_a,
\end{equation}
and inverse transformation
\begin{equation}\label{eq:dX}
\mathrm{d}\mathbf{X}=\mathbf{F}^{-1}\mathrm{d}\mathbf{x},\quad \mathrm{or}\quad \mathrm{d}X_A=F_{Aa}^{-1}\mathrm{d}x_a\mathbf{E}_A,
\end{equation}
where
\begin{equation}\label{eq:defGrad}
\mathbf{F}(\mathbf{X},t)=\frac{\partial \phi(\mathbf{X},t)}{\partial \mathbf{X}},\quad \mathrm{or}\quad F_{aA}=\frac{\partial \phi_a}{\partial X_A}\mathbf{e}_a\varotimes\mathbf{E}_A
\end{equation}
\begin{equation}\label{eq:defGrad}
\mathbf{F}^{-1}(\mathbf{x},t)=\frac{\partial \varphi(\mathbf{x},t)}{\partial \mathbf{x}},\quad \mathrm{or}\quad F_{Aa}^{-1}=\frac{\partial \varphi_A}{\partial x_a}\mathbf{E}_A\varotimes\mathbf{e}_a.
\end{equation}
we postulate the existence of fractional material and spatial line elements as
\begin{equation}\label{eq:dxa}
\mathrm{d}\tilde{\mathbf{x}}=\underset{X}{\tilde{\mathbf{F}}}\mathrm{d}\mathbf{X},\quad \mathrm{or}\quad \mathrm{d}\tilde{x}_a=\underset{X}{\tilde{F}} ~_{aA}\mathrm{d}X_A\mathbf{e}_a,
\end{equation}
and 
\begin{equation}\label{eq:dXa}
\mathrm{d}\tilde{\mathbf{X}}=\underset{x}{\tilde{\mathbf{F}}}\mathrm{d}\mathbf{x},\quad \mathrm{or}\quad \mathrm{d}\tilde{X}_A=\underset{x}{\tilde{F}} ~_{Aa}\mathrm{d}x_a\mathbf{E}_A.
\end{equation}
Using Eqs~(\ref{eq:dx}),(\ref{eq:dX}),(\ref{eq:dxa}) and (\ref{eq:dXa}) we have 
\begin{equation}\label{eq:dxaII}
\mathrm{d}\tilde{\mathbf{x}}=\overset{\alpha}{\mathbf{F}}\mathrm{d}\tilde{\mathbf{X}},\quad \mathrm{or}\quad \mathrm{d}\tilde{x}_a=\overset{\alpha}{F}_{aA}\mathrm{d}\tilde{X}_A\mathbf{e}_a,
\end{equation}

\begin{equation}\label{eq:dXXa}
\mathrm{d}\tilde{\mathbf{X}}=\underset{x}{\overset{\alpha}{\mathbf{F}}}\mathrm{d}\mathbf{X},\quad \mathrm{or}\quad \mathrm{d}\tilde{X}_B=\underset{x}{\overset{\alpha}{{F}}} ~_{BA}\mathrm{d}X_A\mathbf{E}_B,
\end{equation}

\begin{equation}\label{eq:dXXa}
\mathrm{d}\tilde{\mathbf{x}}=\underset{X}{\overset{\alpha}{\mathbf{F}}}\mathrm{d}\mathbf{x},\quad \mathrm{or}\quad \mathrm{d}\tilde{x}_b=\underset{X}{\overset{\alpha}{F}} ~_{ba}\mathrm{d}x_a\mathbf{e}_b,
\end{equation}

where $\overset{\alpha}{\mathbf{F}}=\underset{X}{\tilde{\mathbf{F}}}\mathbf{F}^{-1}\underset{x}{\tilde{\mathbf{F}}} ~^{-1}$, $\underset{x}{\overset{\alpha}{\mathbf{F}}}=\underset{x}{\tilde{\mathbf{F}}}\mathbf{F}$ and $\underset{X}{\overset{\alpha}{\mathbf{F}}}=\underset{X}{\tilde{\mathbf{F}}}\mathbf{F}^{-1}$. Figure~\ref{fig:diagram} summarises the situation.

\begin{myRemark}\label{rem:1}\textbf{On the non-locality of fractional deformation gradients.}
Classical deformation gradient $\mathbf{F}$ is local while fractional ones $\underset{X}{\tilde{\mathbf{F}}}$ and $\underset{x}{\tilde{\mathbf{F}}}$ are non-local due to the definition of RC fractional differential operator. RC is based on an interval whose length depends on a particular material being described. Due to the definitions Eqs~(\ref{eq:fracdefGradX}) and (\ref{eq:fracdefGradx}), their physical/geometrical interpretation, is analogous, compared with the classical one - the  difference is that they operate on fractional or mixed (classical/factional) line elements. 
\end{myRemark}

\begin{myRemark}\label{rem:2}\textbf{On the length scale parameters in fractional deformation gradients definitions.}
The main attention should be paid to the length scale parameters appearing in definitions Eqs~(\ref{eq:fracdefGradX}) and (\ref{eq:fracdefGradx}) of the fractional deformation gradients. 

It is important to notice that without those parameters, the unit of the fractional deformation gradients would be (in SI) $m^{1-\alpha}$. Therefore, the introduction of the length scales, similarly to classical gradient continuum models, allows to finally obtain dimensionless quantity. Therefore, we can compare the lengths of line elements $\mathrm{d}{\mathbf{X}}$ and $\mathrm{d}{\mathbf{x}}$ with fractional ones $\mathrm{d}\tilde{\mathbf{X}}$ and $\mathrm{d}\tilde{\mathbf{x}}$ what is crucial concerning possible strains definitions.

In previous papers, where isotropic non-locality was considered as mentioned, it was shown also, that it is necessary to introduce the length scales in order to fulfil the rigid body motion requirements (objectivity restriction). Nevertheless, from purely mathematical point of view, those parameters could be omitted.
\end{myRemark}

\begin{myRemark}\label{rem:3}\textbf{On the equivalence of the fractional formulation with the classical one for $\alpha_{ij}=1$.}
For $\alpha_{ij}=1$ RC continua becomes classical one and hence we have:
\begin{equation}
\mathbf{F}=\underset{X}{\tilde{\mathbf{F}}}=\underset{x}{\tilde{\mathbf{F}}} ~^{-1}={\overset{\alpha}{\mathbf{F}}},
\end{equation}

\begin{equation}
\mathbf{F}^{-1}=\underset{X}{\tilde{\mathbf{F}}} ~^{-1}=\underset{x}{\tilde{\mathbf{F}}}={\overset{\alpha}{\mathbf{F}}} ~^{-1},
\end{equation}

\begin{equation}
\underset{x}{\overset{\alpha}{\mathbf{F}}}=\mathbf{I},
\end{equation}

\begin{equation}
\underset{X}{\overset{\alpha}{\mathbf{F}}}=\mathbf{i},
\end{equation}

\begin{equation}
\mathrm{d}{\mathbf{x}}=\mathrm{d}\tilde{\mathbf{x}},
\end{equation}

\begin{equation}
\mathrm{d}{\mathbf{X}}=\mathrm{d}\tilde{\mathbf{X}}.
\end{equation}

\end{myRemark}

\begin{myRemark}\label{rem:4}\textbf{On the fractional deformation gradient properties.}
$\underset{x}{\overset{\alpha}{\mathbf{F}}}$ and $\underset{X}{\overset{\alpha}{\mathbf{F}}}$ are not two point tensors while $\underset{X}{\tilde{\mathbf{F}}}$, $\underset{x}{\tilde{\mathbf{F}}}$ and $\overset{\alpha}{\mathbf{F}}$ are. Based on the properties of motion Eq.~(\ref{eq:motion}) we have that the inverse of $\underset{X}{\tilde{\mathbf{F}}}$ and $\underset{x}{\tilde{\mathbf{F}}}$ exists.

\end{myRemark}

\subsection{Fractional strains}\label{sse:frstr}
We define the strains by analogy to the classical continuum mechanics based on the difference in scalar products in actual and reference configurations. The introduced fractional deformation gradients allows to define 4 concepts of strains. Thus, one can define:  

		\begin{equation}\label{eq:Efrac}
			\mathbf{E}=\frac{1}{2}(\overset{\Diamond}{\mathbf{F}} ~^T\overset{\Diamond}{\mathbf{F}}-\mathbf{I}),\quad \mathrm{or}\quad {E}_{AB}=\frac{1}{2}(\overset{\Diamond}{F} ~^T_{Aa}\overset{\Diamond}{F} ~_{aB}-{I}_{AB})\mathbf{E}_A\varotimes\mathbf{E}_B,
\end{equation}

		\begin{equation}\label{eq:efrac}			
		\mathbf{e}=\frac{1}{2}(\mathbf{i}-\overset{\Diamond}{\mathbf{F}} ~^{-T}\overset{\Diamond}{\mathbf{F}} ~^{-1}),\quad \mathrm{or}\quad {e}_{ab}=\frac{1}{2}({i}_{ab}-\overset{\Diamond}{F} ~_{aA}^{-T}\overset{\Diamond}{F} ~^{-1}_{Ab})\mathbf{e}_a\varotimes\mathbf{e}_b,
		\end{equation}

where $\mathbf{E}$ is the classical Green-Lagrange strain tensor or its fractional counterpart; $\mathbf{e}$ is the classical Euler-Almansi strain tensor or its fractional counterpart, and depending on the formulation, $\overset{\Diamond}{\mathbf{F}}$ can be replaced with $\mathbf{F}$ or $\underset{X}{\tilde{\mathbf{F}}}$ or $\underset{x}{\tilde{\mathbf{F}}}$ or $\overset{\alpha}{\mathbf{F}}$. According to the chosen $\overset{\Diamond}{\mathbf{F}}$ the associated others variables like (they all have, to some extent, classical meaning according to Remark \ref{rem:1}, but simultaneously are non-local): the left and right Cauchy-Green tensors; the orthogonal tensor and left or right stretch tensor from polar decomposition of $\overset{\Diamond}{\mathbf{F}}$ can be defined using analogical to classical rules.

\begin{myRemark} {\textbf{On the relation between fractional displacement gradient tensor and fractional strains.}} As in the classical continuum mechanics one can define the relation between strains and displacement gradient tensor utilising introduced fractional gradient tensors $\underset{X}{\tilde{\mathbf{F}}}$ and $\underset{x}{\tilde{\mathbf{F}}}$.

The displacements in the material description $\mathbf{U}$ are defined as:
\begin{equation}
\mathbf{U}(\mathbf{X},t)=\mathbf{x}(\mathbf{X},t)-\mathbf{X},
\end{equation}
and its fractional gradient
\begin{equation}
\mathrm{Grad}\underset{X}{\tilde{\mathbf{U}}}=\underset{X}{\tilde{\mathbf{F}}}-\mathbf{I} ,\quad \mathrm{or}\quad \underset{X}{\ell} ~_{aA}^{\alpha_{aA}-1}\underset{X_A}{D}^{\alpha_{aA}} U_a=(\underset{X}{\tilde{{F}}} ~_{aA} - I_{aA})\mathbf{e}_a\varotimes\mathbf{E}_A,
\end{equation}
thus we have
\begin{equation}\label{eq:fracdefXinU}
\underset{X}{\tilde{\mathbf{F}}}=\mathrm{Grad}\underset{X}{\tilde{\mathbf{U}}}+\mathbf{I}.
\end{equation}

Similarly, the displacements in spatial description $\mathbf{u}$ are defined as
\begin{equation}
\mathbf{u}(\mathbf{x},t)=\mathbf{x}-\mathbf{X}(\mathbf{x},t),
\end{equation}
and its fractional gradient
\begin{equation}
\mathrm{grad}\underset{x}{\tilde{\mathbf{u}}}=\mathbf{i}-\underset{x}{\tilde{\mathbf{F}}},\quad \mathrm{or}\quad \underset{x}{\ell} ~_{Aa}^{\alpha_{Aa}-1}\underset{x_a}{D}^{\alpha_{Aa}} u_A=(i_{Aa}-\underset{x}{\tilde{{F}}} ~_{Aa})\mathbf{E}_A\varotimes\mathbf{e}_a,
\end{equation}
thus we have
\begin{equation}\label{eq:fracdefxinu}
\underset{x}{\tilde{\mathbf{F}}}=\mathbf{i}-\mathrm{grad}\underset{x}{\tilde{\mathbf{u}}}.
\end{equation}

By applying Eqs~(\ref{eq:fracdefXinU}) and (\ref{eq:fracdefxinu}) into the fractional strain definitions Eqs~(\ref{eq:Efrac}) and (\ref{eq:efrac}) we obtain their dependence on the fractional displacement gradients. 

As in classical continuum mechanics, we can introduce small fractional Cauchy strain tensor, we have ($\ell=\underset{X}{\ell}=\underset{x}{\ell}$)
\begin{equation}\label{eq:smallStrain}
\overset{\Diamond}{\bveps}=\frac{1}{2}\left[\mathrm{grad}\underset{x}{\tilde{\mathbf{u}}}
+\mathrm{grad}\underset{x}{\tilde{\mathbf{u}}}^T\right],
\end{equation}
where $\overset{\Diamond}{\bveps}$ stands for fractional Cauchy strain. For 1D deformation  Eq.~(\ref{eq:smallStrain}) reads
\begin{equation}\label{eq:smallStr1D-2}
\overset{\Diamond}{\varepsilon} =\frac{\varsigma(\alpha)}{2}\ell^{\alpha-1}
\left(_{x-a}^{~~~~~C}D_x^{\alpha} u- _{x}^{C}D^{\alpha}_{x+b} u\right).
\end{equation}
Notice, that Eq.~(\ref{eq:smallStr1D-2}) is similar to definitions presented in \cite{Klimek2001} (Eq.~9), \cite{Lazopoulos2006}, \cite{Atanackovic2009} (Eq.~2.7) and \cite{Carpinteri2011} (Eq.~20), but operates on finite interval (similarly to the 'short memory' principle discussed in \cite{Podlubny1999-AP}), and gives non-dimensional quantity. It should be emphasised, that in contrast to isotropic non-locality discussed in previous papers (cf. \cite{Sumelka2013-TH}) $a\neq b$, but it remains physically reasonable to assume some relation between $\ell$ and length of internal over which fractional derivative is calculates e.g. equivalence.

\end{myRemark}

\subsection{Stresses and balance laws}

According to Fig.~\ref{fig:diagram}, material line element is $\mathrm{d}{{\mathbf{X}}}$ or $\mathrm{d}{\tilde{\mathbf{X}}}$, while spatial $\mathrm{d}{{\mathbf{x}}}$ or $\mathrm{d}{\tilde{\mathbf{x}}}$ (together with corresponding fractional/classical deformation gradient). The selection of a specific path causes that the remaining one can be considered utilising the concept of dual variables \cite{Haupt-2002}. For example, assuming that we describe the deformation in terms of $\mathrm{d}{{\mathbf{X}}}$ and $\mathrm{d}{\tilde{\mathbf{x}}}$, classical deformation (based on $\mathrm{d}{{\mathbf{X}}}$ and $\mathrm{d}{{\mathbf{x}}}$) appears as (one of possible) 'intermediate' one. Therefore, it is clear that stresses, and balance laws should be postulated dependently on chosen description of deformation in a standard manner (such result is analogous to the one presented in \cite{Atanackovic2009} Eqs 2.7,2.17, and 2.18).

To understand this logic (without loss of generality) let us consider purely mechanical problem. It will be observed that the restrictions to be held by fractional stresses are analogous to the classical one, but fulfilled in auxiliary 'fractional' (or phenomenological) space.  

First, following Fig.~\ref{fig:diagram}, the material (spatial) volume element $\mathrm{d}V$ ($\mathrm{d}v$) or its fractional counterpart $\mathrm{d}\tilde{V}$ ($\mathrm{d}\tilde{v}$), and the material (spatial) surface element $\mathrm{d}\mathbf{S}$ ($\mathrm{d}\mathbf{s}$) or its fractional counterpart $\mathrm{d}\tilde{\mathbf{S}}$ ($\mathrm{d}\tilde{\mathbf{s}}$) are introduced - cf. \cite{Sumelka2014-AoM}. Next, the fractional Cauchy (true) traction vector $\tilde{\mathbf{{t}}}_{(\tilde{\mathbf{n}})}$ exerted on $d\tilde{s}$ with outward normal $\tilde{\mathbf{{n}}}$ is obtained as a transformation of classical Cauchy (true) traction vector $\mathbf{{t}_{({n})}}$ exerted on $d{s}$ with outward normal ${\mathbf{{n}}}$. Thus, according to the diagram in Fig.~\ref{fig:diagram} we have
\begin{equation}
\tilde{\mathbf{{t}}}_{(\tilde{\mathbf{n}})}=\underset{X}{\overset{\alpha}{\mathbf{F}}}\mathbf{{t}_{({n})}}
\quad\textrm{or}\quad
{\tilde{t}_{(\tilde{n})}} ~_b=\underset{X}{\overset{\alpha}{{F}}} ~_{ba}{{t}_{({n})}} ~_a\mathbf{e}_b.
\end{equation}
As in classical set-up we postulate the relationship between fractional traction and fractional Cauchy stress tensor
\begin{equation}
\tilde{\mathbf{{t}}}_{(\tilde{\mathbf{n}})}=\tilde{\mathbf{{n}}}\tilde{\bsigma},
\end{equation}
where $\tilde{\bsigma}$ denotes fractional Cauchy stress tensor. Based on the above relations, and the fact that 
\begin{equation}
\tilde{\mathbf{{n}}}=\underset{X}{\overset{\alpha}{\mathbf{F}}}\mathbf{{n}},
\end{equation}
we have
\begin{equation}
\underset{X}{\overset{\alpha}{\mathbf{F}}}\mathbf{{t}_{({n})}}=
\underset{X}{\overset{\alpha}{\mathbf{F}}}\mathbf{{n}}{\bsigma}=
\underset{X}{\overset{\alpha}{\mathbf{F}}}\mathbf{{n}}\tilde{\bsigma},
\end{equation}
thus finally
\begin{equation}\label{eq:sigmasigma}
\tilde{\bsigma}={\bsigma},
\end{equation}
where ${\bsigma}$ is classical Cauchy stress tensor. Thus, according to relation Eq.~(\ref{eq:sigmasigma}) the fractional stresses are transformation of the classical Cauchy measure to auxiliary 'fractional' space.

Finally, before the balance of momentum in spatial description is postulated for fractional case, the conservation of mass is considered in a form
\begin{equation}
{\rho_0}\mathrm{d}{V}=\tilde{\rho_0}\mathrm{d}\tilde{V}={\rho}\mathrm{d}{v}=\tilde{\rho}\mathrm{d}\tilde{v},
\end{equation}
or shortly
\begin{equation}
 \overset{\Diamond}{\rho_0}=\overset{\Diamond}{J}\overset{\Diamond}{\rho},
\end{equation}
where $\rho_0$ ($\tilde{\rho_0}$) is the reference mass density (fractional counterpart), $\rho$ ($\tilde{\rho}$) is spatial mass density (fractional counterpart), and $\overset{\Diamond}{J}=\mathrm{det}\overset{\Diamond}{\mathbf{F}}$ is a Jacobian (as before in the denotation $\overset{\Diamond}{(\cdot)}$ one can replace $(\cdot)$ with classical quantity or fractional counterpart).
The following relation holds
\begin{equation}\label{eq:ZZP}
\int_{\overset{\Diamond}{v}}\overset{\Diamond}{\rho}\overset{\Diamond}{\dot{\bup}}\mathrm{d}\overset{\Diamond}{v}=\int_{\partial\overset{\Diamond}{v}}\overset{\Diamond}{\mathbf{t}}\mathrm{d}\overset{\Diamond}{{s}}+\int_{\overset{\Diamond}{v}}{\overset{\Diamond}{\rho}\mathbf{f}}\mathrm{d}\overset{\Diamond}{{v}},
\end{equation}
where $\bup$ is a velocity, and $\mathbf{f}$ is a body force per unit mass.
By applying the divergence theorem to Eq.~(\ref{eq:ZZP}) we have 
\begin{equation}
\int_{\overset{\Diamond}{v}}\overset{\Diamond}{\rho}\overset{\Diamond}{\dot{\bup}}\mathrm{d}\overset{\Diamond}{v}=\int_{\overset{\Diamond}{v}}\mathrm{div}\overset{\Diamond}{\bsigma} ~^{T}\mathrm{d}\overset{\Diamond}{{v}}+\int_{\overset{\Diamond}{v}}{\overset{\Diamond}{\rho}\mathbf{f}}\mathrm{d}\overset{\Diamond}{{v}},
\end{equation}
so 
\begin{equation}\label{eq:local}
\mathrm{div}\overset{\Diamond}{\bsigma} ~^{T}+{\overset{\Diamond}{\rho}\mathbf{f}}=\overset{\Diamond}{\rho}\overset{\Diamond}{\dot{\bup}},
\end{equation}
or in the absence of inertia forces
\begin{equation}
\mathrm{div}\overset{\Diamond}{\bsigma} ~^{T}+{\overset{\Diamond}{\rho}\mathbf{f}}=\mathbf{0}.
\end{equation}
Furthermore, the symmetry of $\overset{\Diamond}{\bsigma}$ (so $\bsigma$ or $\tilde{\bsigma}$) can be checked in classical manner utilising the balance of moment of momentum.

As a concluding remark, it is clear that formulation in a material description needs proper definitions for the first and the second (fractional) Piola-Kirchhoff stress tensors. Thus, dependently on chosen fractional strains definition we have:

\begin{equation}
{\mathbf{P}}=J\bsigma{\mathbf{F}^{-T}},\quad{\mathbf{S}}=J{\mathbf{F}^{-1}}\bsigma{\mathbf{F}^{-T}}={\mathbf{F}^{-1}}{\mathbf{P}},
\end{equation}

\begin{equation}
\underset{X}{\tilde{\mathbf{P}}}=\underset{X}{\tilde{J}}\bsigma{\underset{X}{\tilde{\mathbf{F}}}}^{-T},\quad{\underset{X}{\tilde{\mathbf{S}}}}=
\underset{X}{\tilde{J}}{\underset{X}{\tilde{\mathbf{F}}}}^{-1}\bsigma{\underset{X}{\tilde{\mathbf{F}}}}^{-T}={\underset{X}{\tilde{\mathbf{F}}}}^{-1}{\underset{X}{\tilde{\mathbf{P}}}},
\end{equation}

\begin{equation}
\underset{x}{\tilde{\mathbf{P}}}={\underset{x}{\tilde{J}}}^{-1}\bsigma{\underset{x}{\tilde{\mathbf{F}}}}^{-T},\quad{\underset{x}{\tilde{\mathbf{S}}}}=
{\underset{x}{\tilde{J}}}^{-1}{\underset{x}{\tilde{\mathbf{F}}}}\bsigma{\underset{x}{\tilde{\mathbf{F}}}}^{-T}={\underset{x}{\tilde{\mathbf{F}}}}{\underset{x}{\tilde{\mathbf{P}}}},
\end{equation}

\begin{equation}
\overset{\alpha}{\mathbf{P}}=\overset{\alpha}{J}\bsigma{\overset{\alpha}{\mathbf{F}}}^{-T},\quad{\overset{\alpha}{\mathbf{S}}}=
\overset{\alpha}{J}{\overset{\alpha}{\mathbf{F}}}^{-1}\bsigma{\overset{\alpha}{\mathbf{F}}}^{-T}={\overset{\alpha}{\mathbf{F}}}^{-1}{\overset{\alpha}{\mathbf{P}}},
\end{equation}

where ${\mathbf{P}},\underset{X}{\tilde{\mathbf{P}}},\underset{x}{\tilde{\mathbf{P}}},\overset{\alpha}{\mathbf{P}}$ and ${\mathbf{S}},\underset{X}{\tilde{\mathbf{S}}},\underset{x}{\tilde{\mathbf{S}}},\overset{\alpha}{\mathbf{S}}$  are classical/fractional the first and the second Piola-Kirchhoff stress tensors, respectively. As previously, taking $\alpha_{ij}=1$ (local kinematics) the introduced stresses definitions reduce to the corresponding classical one
\begin{equation}
{\mathbf{P}}=\underset{X}{\tilde{\mathbf{P}}}=\underset{x}{\tilde{\mathbf{P}}}=\overset{\alpha}{\mathbf{P}},
\end{equation}
\begin{equation}
{\mathbf{S}}=\underset{X}{\tilde{\mathbf{S}}}=\underset{x}{\tilde{\mathbf{S}}}=\overset{\alpha}{\mathbf{S}}.
\end{equation}

\begin{myRemark} {\textbf{On physical interpretation of the parameters $\alpha$ and $\ell$.}} Introducing non-locality utilising fractional calculus we simultaneously add a set material parameters $\alpha$ and $\ell$. The order of fractional continua $\alpha$ (together with type of applied fractional differential operator) controls the way in which the information from the surrounding influences particular point of interest. The length scale $\ell$ defines the amount of this information (size of this surrounding). So, both parameters are crucial. In the following example we will observe that both $\alpha\shortrightarrow 1$ and $\ell\shortrightarrow 0$ recovers classical solution - as it should be.

\end{myRemark}

\section{Examples}
In the example section, the influence of fractional kinematics on the fractional strains is presented based on the assumed different motions. The following notations for strains are proposed (cf. Eqs~(\ref{eq:Efrac}) and (\ref{eq:efrac})):

\begin{enumerate}
    \item Classical formulation
		\begin{equation}
		\mathrm{E}=\frac{1}{2}(\mathbf{F}^T\mathbf{F}-\mathbf{I}), \quad
\mathbf{e}=\frac{1}{2}(\mathbf{i}-\mathbf{F}^{-T}\mathbf{F}^{-1}).
		\end{equation}

	\item Formulation based on the fractional spatial line element ($\mathrm{d}{\tilde{\mathbf{x}}}$) and the classical material line element $(\mathrm{d}{\mathbf{X}})$

		\begin{equation}
		\underset{X}{\tilde{\mathbf{E}}}=\frac{1}{2}(\underset{X}{\tilde{\mathbf{F}}} ~^T\underset{X}{\tilde{\mathbf{F}}}-\mathbf{I}), \quad
\underset{X}{\tilde{\mathbf{e}}}=\frac{1}{2}(\mathbf{i}-\underset{X}{\tilde{\mathbf{F}}} ~^{-T}\underset{X}{\tilde{\mathbf{F}}} ~^{-1}).
		\end{equation}

	\item Formulation based on the classical spatial line element ($\mathrm{d}{{\mathbf{x}}}$) and the fractional material line element $(\mathrm{d}{\tilde{\mathbf{X}}})$

		\begin{equation}
		\underset{x}{\tilde{\mathbf{E}}}=\frac{1}{2}(\underset{x}{\tilde{\mathbf{F}}} ~^{-T}\underset{x}{\tilde{\mathbf{F}}} ~^{-1}-\mathbf{I}), \quad
\underset{x}{\tilde{\mathbf{e}}}=\frac{1}{2}(\mathbf{i}-\underset{x}{\tilde{\mathbf{F}}} ~^{T}\underset{x}{\tilde{\mathbf{F}}} ).
		\end{equation}

	\item Formulation based on the fractional spatial line element ($\mathrm{d}{\tilde{\mathbf{x}}}$) and the fractional material line element $(\mathrm{d}{\tilde{\mathbf{X}}})$

		\begin{equation}
		\overset{\alpha}{\mathbf{E}}=\frac{1}{2}(\overset{\alpha}{\mathbf{F}} ~^T{\overset{\alpha}{\mathbf{F}}}-\mathbf{I}), \quad
\overset{\alpha}{\mathbf{e}}=\frac{1}{2}(\mathbf{i}-\overset{\alpha}{\mathbf{F}} ~^{-T}\overset{\alpha}{\mathbf{F}} ~^{-1}).
		\end{equation}
	
\end{enumerate} 

In following examples 1D deformation is considered. We assume for brevity that: (i) $\ell=\underset{X}{\ell}=\underset{x}{\ell}$; (ii) $0<\alpha\leq 1$; (iii) the deformation in a particular point of interest $X_A$ is calculated for terminals $a=X_A-\ell_L$ and $b=\ell_R+X_A$ (this is because of anisotropy of non-locality that $X_A$ in general does not necessarily lay in the middle of interval $(a,b)$). 

\subsection{Example 1} Let us consider a particular type of motion, namely the case when motion is a linear function of coordinate. We compare for brevity the deformation based on $\mathbf{F}$ and $\underset{X}{\tilde{\mathbf{F}}}$. Thus, the motion is described as 
\begin{equation}
\mathbf{x}=\phi(\mathbf{X})=(1+\beta)X_1\mathbf{e}_1+X_2\mathbf{e}_2+X_3\mathbf{e}_3.
\end{equation}

For such a case we have
\begin{equation}
\mathbf{F} = 
\begin{bmatrix} 
(1+\beta) & 0 & 0 \\
0 & 1 & 0 \\
0 & 0 & 1
\end{bmatrix},
\end{equation}
and (we omit the underset $X$ for $\varsigma$)
\begin{equation}
\underset{X}{\tilde{\mathbf{F}}} = \mathcal{M}
\begin{bmatrix} 
(1+\beta) & 0 & 0 \\
0 & 1 & 0 \\
0 & 0 & 1
\end{bmatrix},
\end{equation}
thus
\begin{equation}
\mathbf{E} =\frac{1}{2}
\begin{bmatrix} 
(1+\beta)^2-1 & 0 & 0 \\
0 & 0 & 0 \\
0 & 0 & 0
\end{bmatrix},
\end{equation}
and
\begin{equation}
\underset{X}{\tilde{\mathbf{E}}} = \frac{1}{2}
\begin{bmatrix} 
[\mathcal{M}(1+\beta)]^2-1 & 0 & 0 \\
0 & \mathcal{M}^2-1 & 0 \\
0 & 0 & \mathcal{M}^2-1
\end{bmatrix}
\end{equation}

where $\mathcal{M}=\frac{\varsigma(\alpha)}{2\Gamma(2-\alpha)}\ell^{\alpha-1}\left(\ell_L^{1-\alpha}+\ell_R^{1-\alpha}\right)$. We observe, that anisotropic fractional deformation introduces in general directional deformation e.g. due to evolving micro-structure or existence of electro-magnetic, thermic, and/or chemical processes \cite{Drapaca2012}.

\subsection{Example 2} Lets consider a motion being a non-linear function of the material coordinates $X_A$. As before, for brevity, the deformation based on $\mathbf{F}$ and $\underset{X}{\tilde{\mathbf{F}}}$ is compared. Thus, the motion is described by 
\begin{equation}
\mathbf{x}=\phi(\mathbf{X})=e^{X_1}\mathbf{e}_1+X_2\mathbf{e}_2+X_3\mathbf{e}_3.
\end{equation}

In this case the classical solution is presented as
\begin{equation}
\mathbf{F} = 
\begin{bmatrix} 
e^{X_1} & 0 & 0 \\
0 & 1 & 0 \\
0 & 0 & 1
\end{bmatrix} \Rightarrow
\mathbf{E} = \frac{1}{2}
\begin{bmatrix} 
e^{2X_1}-1 & 0 & 0 \\
0 & 0 & 0 \\
0 & 0 & 0
\end{bmatrix}.
\end{equation}

The calculation of $\underset{X}{\tilde{\mathbf{F}}}$ requires detailed explanation. By analogy to the first example $\underset{X}{\tilde{{F}}} ~_{22}=\underset{X}{\tilde{{F}}} ~_{33}=\mathcal{M}$ whereas  $\underset{X}{\tilde{{F}}} ~_{11}$ will be approximated numerically, because analytical solutions under fractional calculus are very limited \cite{Odibat2006,Leszczynski2011}. Therefore, according to Eq.~(\ref{eq:RC}) we need to calculate adequate left and right Caputo derivatives for $\phi$. We have
\begin{equation}
\underset{X}{\tilde{{F}}} ~_{11} =\ell^{\alpha-1} ~_{~~a}^{RC}\underset{X_1}{D} ~^{\alpha}_b e^{X_1}=\ell^{\alpha-1} \frac{\varsigma(\alpha)}{2} \left(~_{X_1-\ell_L}^{~~~~~~~~C}D^{\alpha}_{X_1} e^{X_1} - ~_{X_1}^{~~C}D^{\alpha}_{X_1+\ell_R}  e^{X_1}\right).
\end{equation}
where for calculations purposes $\varsigma(\alpha)=\Gamma(2-\alpha)$, and $\ell=\frac{1}{2}(\ell_L+\ell_R)$.

The following approximations utilising the modified trapezoidal rule can be used for the calculations \cite{Leszczynski2011}. For the left sided derivatives we use:
\begin{equation}
a=t_0<t_1<...<t_j<...<t_m=t, \quad h=\frac{t_m-t_0}{m}=\frac{t-a}{m}, \quad m \geq 2,
\end{equation}
\begin{eqnarray}
\nonumber
~_{a}^{C}D^{\alpha}_{t} f(t)\cong\frac{h^{n-\alpha}}{\Gamma(n-\alpha+2)}\left\{[(m-1)^{n-\alpha+1}-(m-n+\alpha-1)m^{n-\alpha}]f^{(n)}(t_0)\right.+\\
f^{(n)}(t_m)+\sum_{j=1}^{m-1}[(m-j+1)^{n-\alpha+1}-2(m-j)^{n-\alpha+1}+\left.(m-j-1)^{n-\alpha+1}]f^{(n)}(t_j)\right\},
\end{eqnarray}

where $f^{(n)}(t_j)$ denotes classical $n$-th derivative at $t=t_j$.

Similarly, for the right sided derivatives we use:
\begin{equation}
t=t_0<t_1<...<t_j<...<t_m=b, \quad h=\frac{t_m-t_0}{m}=\frac{b-t}{m}, \quad m \geq 2,
\end{equation}
\begin{eqnarray}
\nonumber
~_{t}^{C}D^{\alpha}_{b} f(t)\cong (-1)^n\frac{h^{n-\alpha}}{\Gamma(n-\alpha+2)}\left\{[(m-1)^{n-\alpha+1}-(m-n+\alpha-1)m^{n-\alpha}]f^{(n)}(t_m)\right.+\\
f^{(n)}(t_0)+\sum_{j=1}^{m-1}[(j+1)^{n-\alpha+1}-2j^{n-\alpha+1}+\left.(j-1)^{n-\alpha+1}]f^{(n)}(t_j)\right\}.
\end{eqnarray}

\begin{figure}[H]
\centering
\includegraphics[width=13cm]{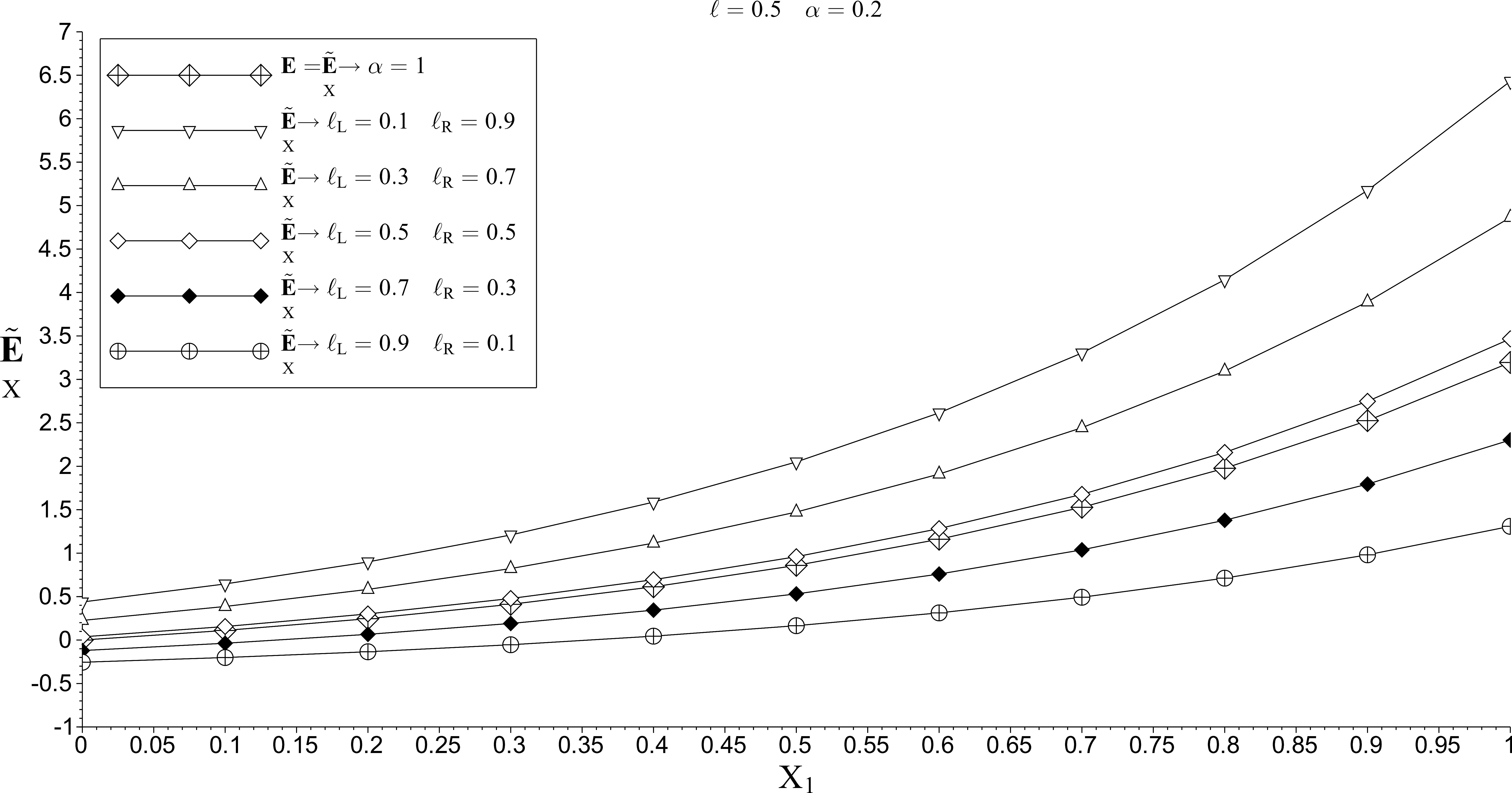}
\includegraphics[width=13cm]{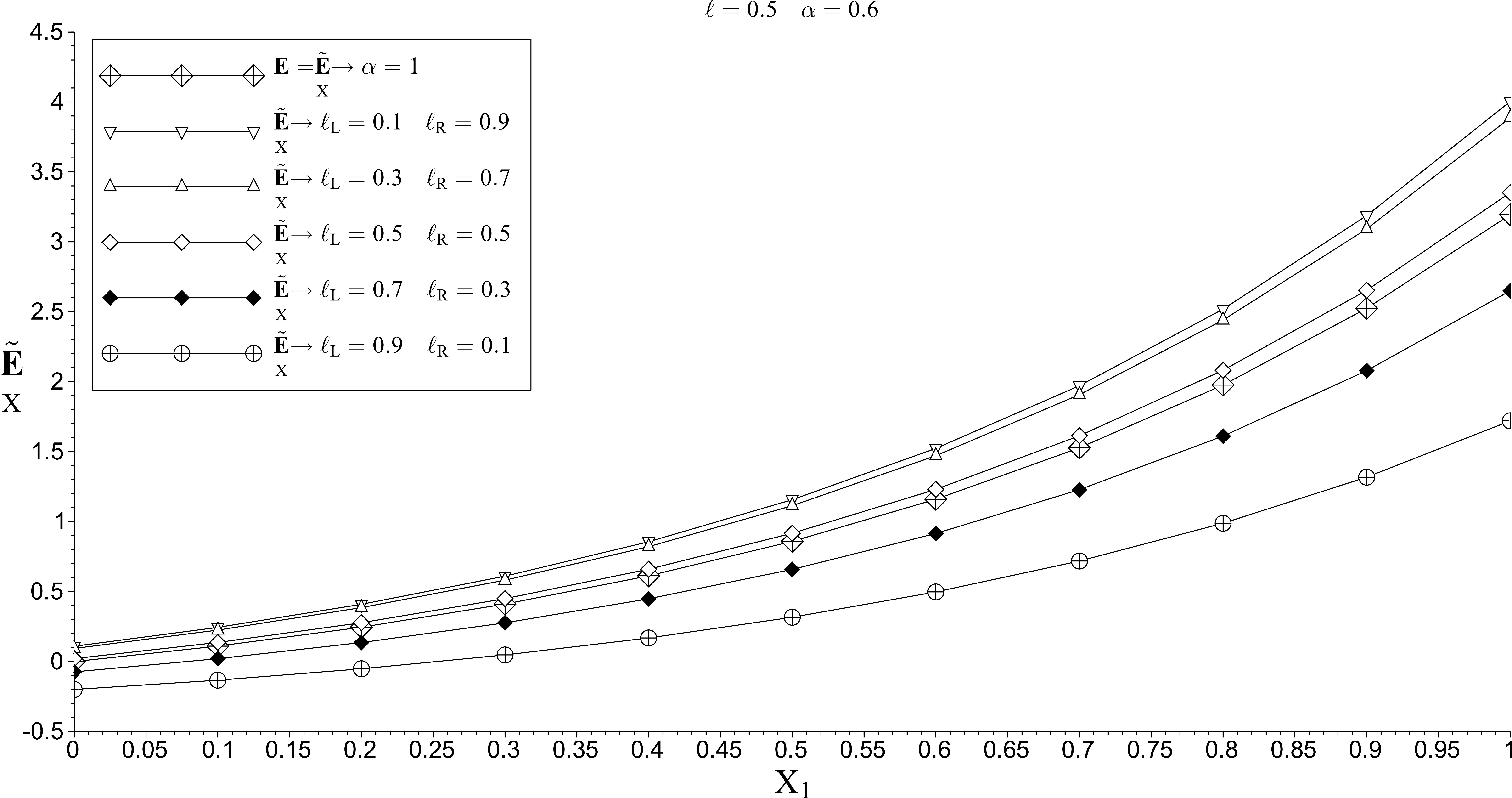}
\includegraphics[width=13cm]{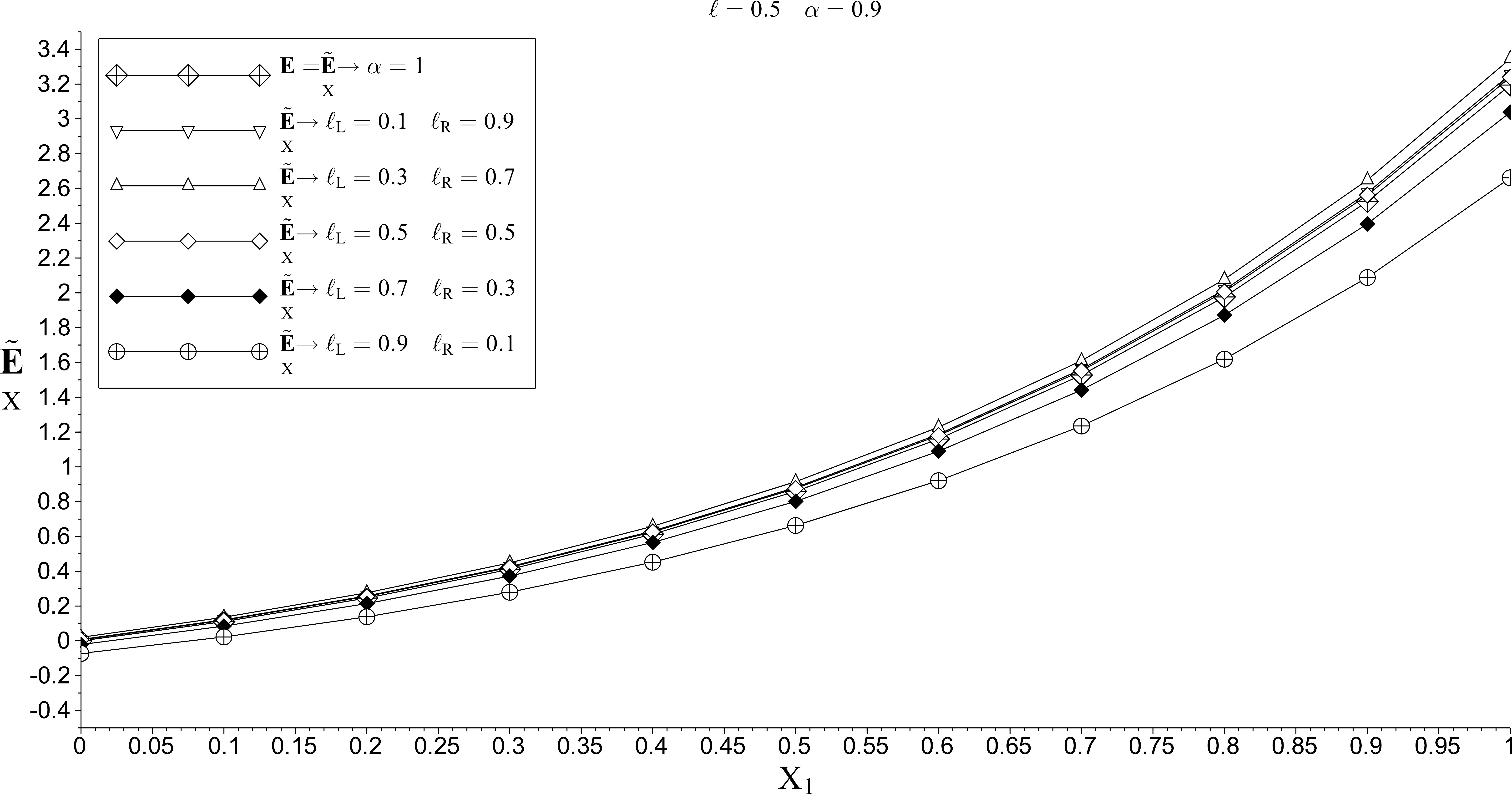}
\caption{The fractional strains $\underset{X}{\tilde{\mathbf{E}}}$ and $\underset{X}{\tilde{\mathbf{e}}}$ against the anisotropy of non-locality and the order of derivative $\alpha$ for $\ell=0.5$}
\label{fig:E2}
\end{figure}

\begin{figure}[H]
\centering
\includegraphics[width=13cm]{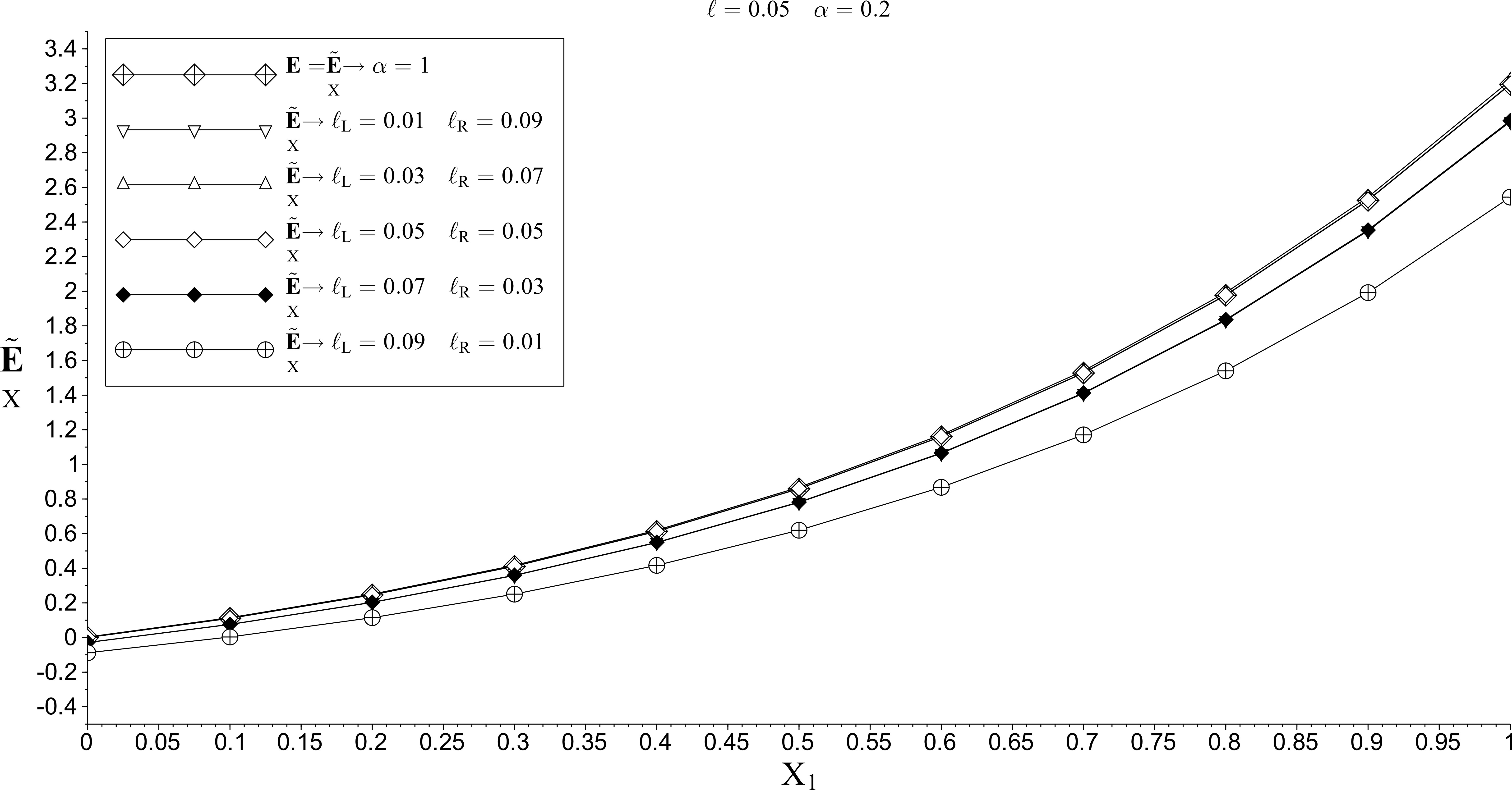}
\includegraphics[width=13cm]{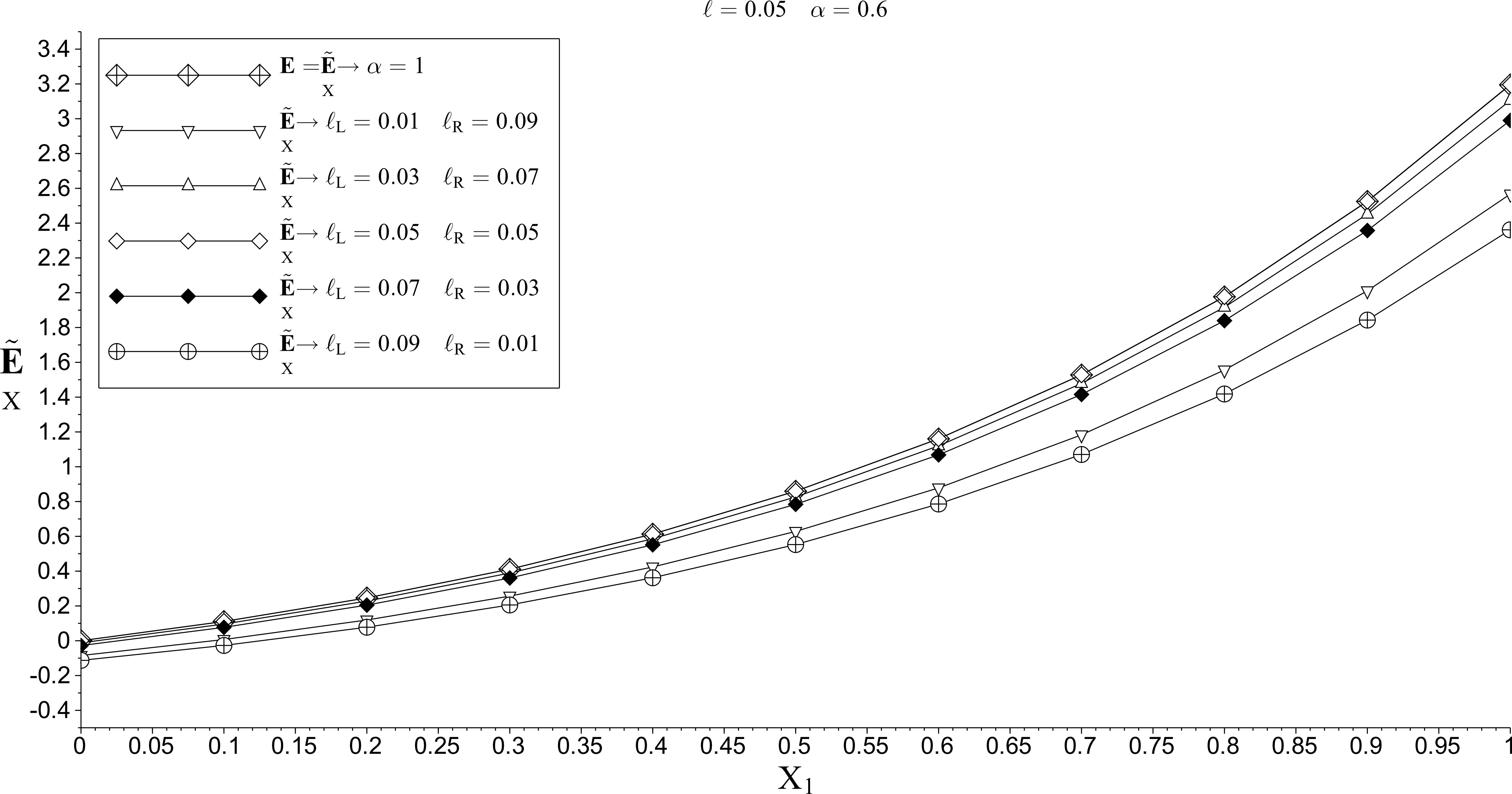}
\includegraphics[width=13cm]{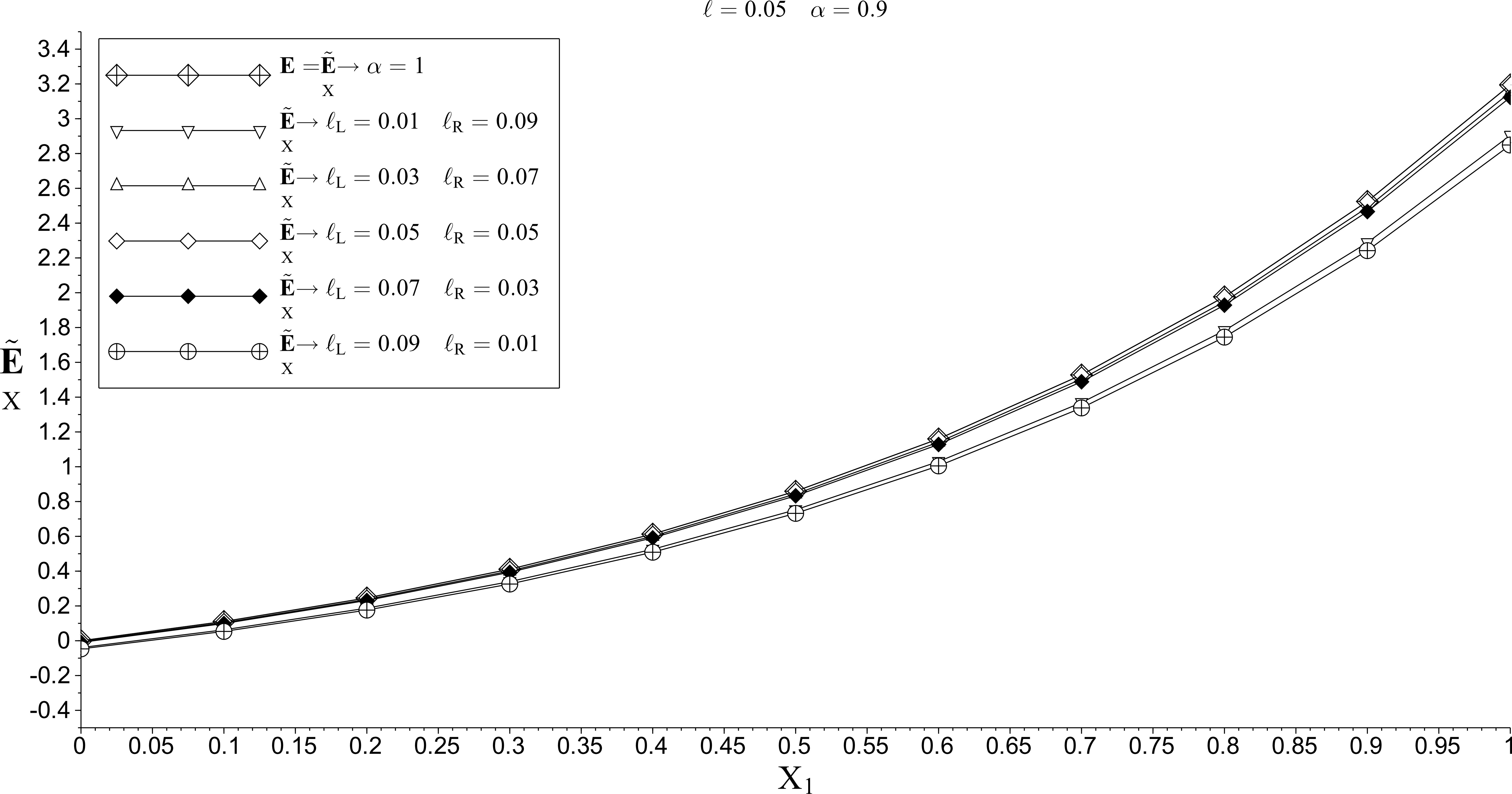}
\caption{The fractional strains $\underset{X}{\tilde{\mathbf{E}}}$ and $\underset{X}{\tilde{\mathbf{e}}}$ against the anisotropy of non-locality and the order of derivative $\alpha$ for $\ell=0.05$}
\label{fig:E3}
\end{figure}

\begin{figure}[H]
\centering
\includegraphics[width=13cm]{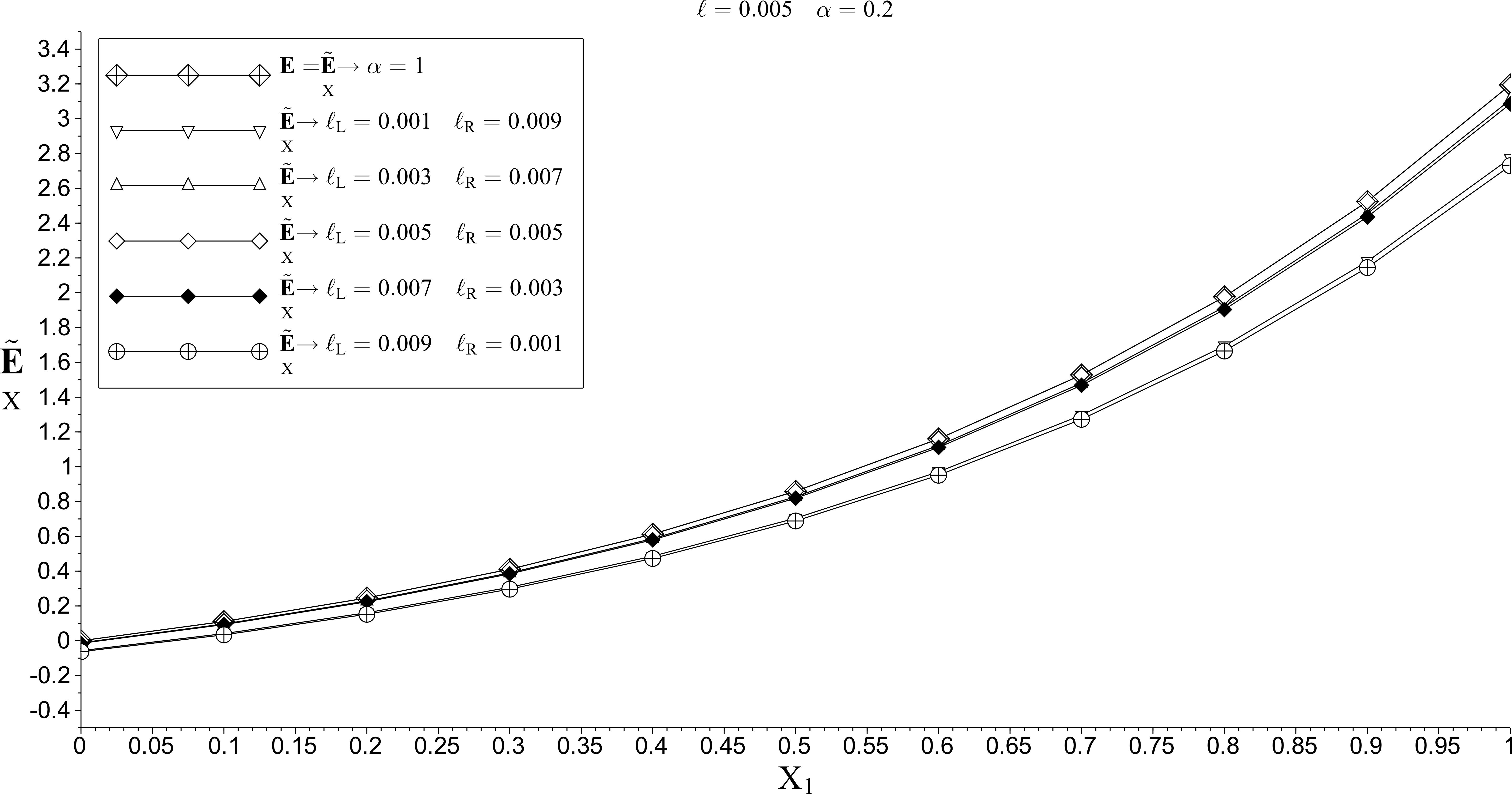}
\includegraphics[width=13cm]{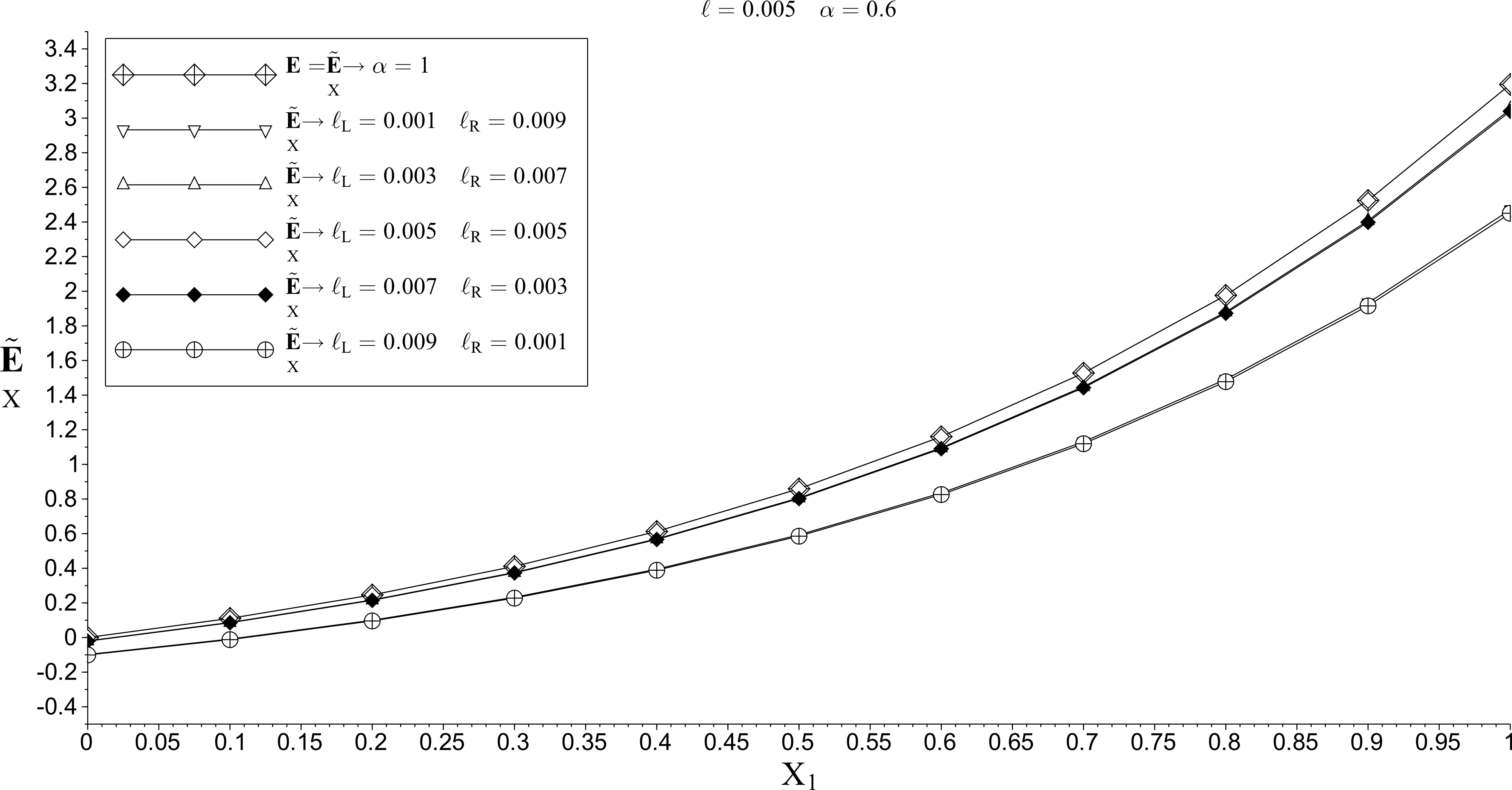}
\includegraphics[width=13cm]{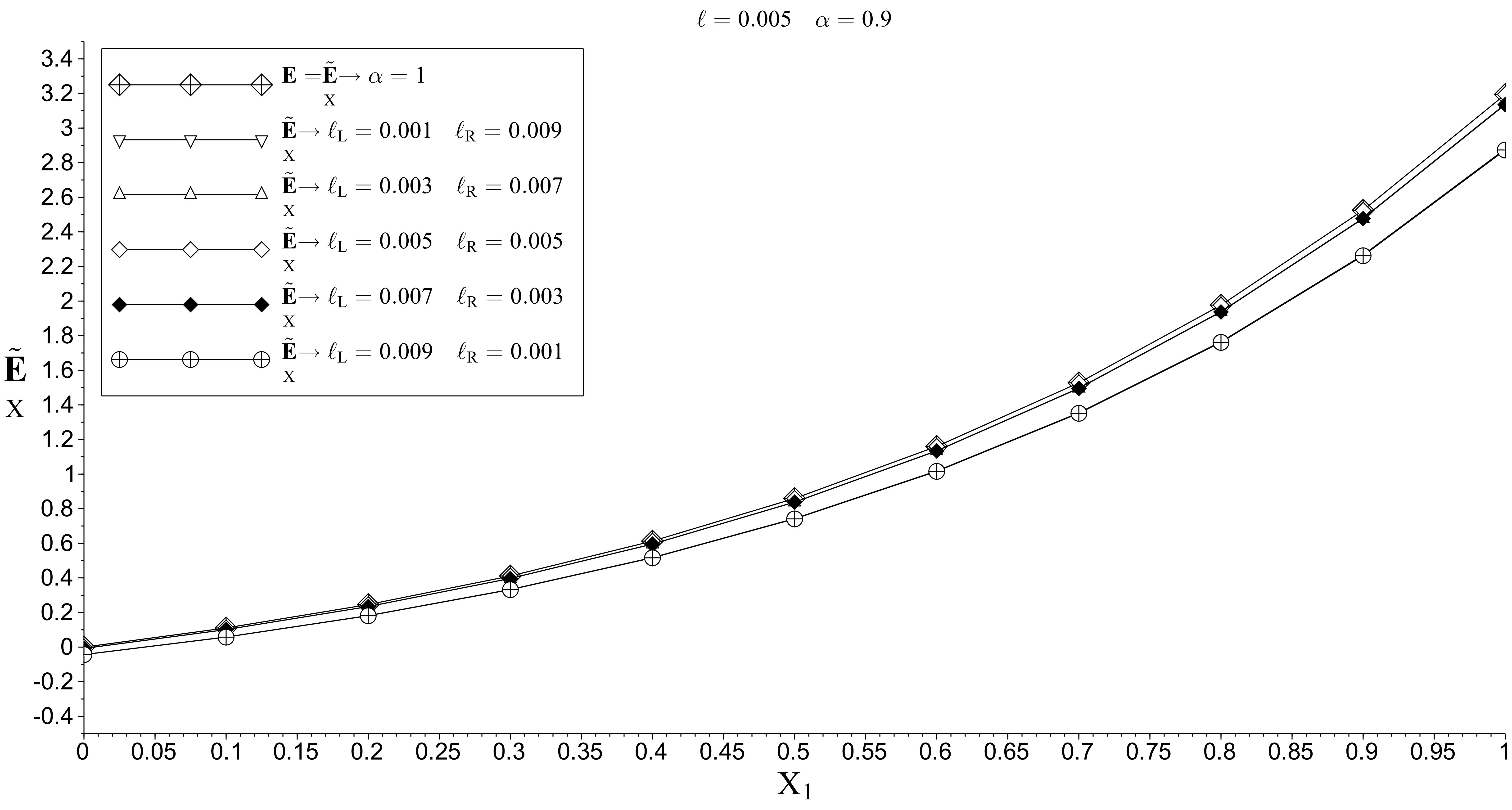}
\caption{The fractional strains $\underset{X}{\tilde{\mathbf{E}}}$ and $\underset{X}{\tilde{\mathbf{e}}}$ against the anisotropy of non-locality and the order of derivative $\alpha$ for $\ell=0.005$}
\label{fig:E4}
\end{figure}

The obtained results are shown in Figs~(\ref{fig:E2}),~(\ref{fig:E3})~and~(\ref{fig:E4}), where the comparison of fractional strain measure $\underset{X}{\tilde{\mathbf{E}}}$ against the anisotropy of non-locality, the order of derivative $\alpha$, and length scale $\ell$ is presented.

It should be noticed that for $\ell\rightarrow 0$ (thus, simultaneously the length of the interval over which fractional measures are calculated approaches to zero) fractional strain measure merge with the classical definition (which are defined in a point) independently of the chosen order of fractional derivative and anisotropy of non-locality. Similarly, if $\alpha\rightarrow 1$ once more the classical strains are captured, however independently of chosen $\ell$.

Another interesting result is the observation that one could model size effect utilising fractional strains. It is clearly presented in Figs~(\ref{fig:E2}),~(\ref{fig:E3})~and~(\ref{fig:E4}) where the dependence of the results against assumed length scale $\ell$ is shown.

Recall, that the introduced fractional strains measures have meaning analogous to the classical one. The difference is that the particular value of strains in a point includes the information from its finite surrounding (controlled by $\ell_L$ and $\ell_R$) in the way defined by order of fractional continua (controlled by $\alpha$).

\section{Conclusions}

In this paper the new concept of generalisation of the classical continuum mechanics utilising fractional calculus is presented. Due to fractional derivative properties the obtained formulation is \textit{non-local}, and this non-locality is of \textit{anisotropic} type.  It should be pointed out that the defined non-local model can be applied to any known constitutive model of elasticity, hyper/hypo-elasticity, visco/plasticity etc. 

The \textit{anisotropic} non-locality introduces a set material parameters: {orders of fractional continua ($\alpha_{ij}$) and length scales ($\ell_{ij}$)}. They  control the way in which the information from the surrounding influences particular point of interest. This result by far enhances modelling approach and allows closely mimic the experimental observations.

Based on two examples picturing one dimensional deformation, the role of the non-local definition of strains based on the fractional differential operators is presented along with the numerical treatment of the fractional derivatives. Concluding, fractional models of mechanics give the opportunity for deeper insight into the analysed problem contrary to classical local formulation.

\section*{Acknowledgment}
This work is supported by the National Centre for Research and Development (NCBiR) under Grant No. UOD-DEM-1-203/001.

\end{document}